\documentclass[%
 sor,
 jor,
 amsmath,amssymb,
 reprint, floatfix,%
author-numerical
]{revtex4-1}

\usepackage{graphicx}
\usepackage{dcolumn}
\usepackage{bm}

\usepackage{epstopdf}
\usepackage{cmap}
\usepackage[english]{babel}
\usepackage{amsmath,amssymb,amsfonts}
\usepackage{hyperref}

\usepackage{url}
\usepackage[hyphenbreaks]{breakurl}

\usepackage{subfigure}

\hypersetup{colorlinks=true,unicode,pdfhighlight=/P}

\begin{document}


\title[Nonlinear resonances in the $ABC$-flow]{Nonlinear resonances in the $ABC$-flow}

\author{A. A. Didov}
 \email{Kedr\_ad@mail.ru}
\affiliation{School of Natural Sciences, Far Eastern Federal University, 690090 Vladivostok, Russia}%
\affiliation{Laboratory of Nonlinear Dynamical Systems, Pacific Oceanological Institute of the Russian Academy of Sciences, 690041 Vladivostok, Russia}

\author{M. Yu. Uleysky}
 \email{uleysky@poi.dvo.ru}
\affiliation{Laboratory of Nonlinear Dynamical Systems, Pacific Oceanological Institute of the Russian Academy of Sciences, 690041 Vladivostok, Russia}

\date{\today}

\begin{abstract}
In this paper we study resonances of the $ABC$-flow in the near integrable case ($C\ll 1$). This is an interesting example of a Hamiltonian system with 3/2 degrees
of freedom in which simultaneous existence of two resonances of the same order is possible.
Analytical conditions of the resonance existence are received. It is shown numerically that
the largest $n:1$ ($n=1,2,3$) resonances exist, and their energies are equal to theoretical
energies in the near integrable case. We provide analytical and numerical evidences
for existence of two branches of the two largest $n:1$ ($n=1,2$) resonances in the region of finite motion.
\end{abstract}


\maketitle

\begin{quotation}

\bf{The $ABC$-flow is a 3D simple model for studying different nonlinear dynamical
processes, e.g., in astrophysics of magnetic fields in stars and in hydrodynamical flows.
We investigate the influence of internal properties of the system on
appearance of new structures and propose a method to describe that. The system has only three control parameters. Assuming one of them to be zero, we get two integrable subsystems
with 2D motion in a horizontal plane and 1D vertical motion. If one of the
parameters is close to zero, the periodic vertical motion can be considered as an external perturbation to the motion in a horizontal plane. As the result, the frequency of external perturbation in the system depends on its Hamiltonian, which allows us to observe the existence of several chains of resonances of one order and reconnection of their separatrices under certain
conditions.}
\end{quotation}

\section{Introduction}
The well-known Arnold--Beltrami--Childress ($ABC$) flow is a steady solution of the Euler equations.
Furthermore, the $ABC$-flow can be considered  as a solution of the Navier-Stokes equations.

The $ABC$-flow is known in fluid mechanics, and it has been studied by many authors.
It may be considered as a simple example of a inviscid fluid flow
with intense mixing associated with substantial helicity. This Lagrangian complexity and the
Beltrami property ($\left[\vec{V} , \left[\nabla,\vec{V}\right]\right]=0$)
make it of great interest in both hydrodynamics and magnetohydrodynamics.
The $ABC$-flow is a prototype for the fast dynamo action, essential to the origin of magnetic field
for large astrophysical objects. Exponential stretching of fluid elements, which is typical for chaotic systems, is necessary for fast dynamo action \cite{Zeldovich1972,Vishik1989}.

The $ABC$-flow was firstly introduced by V. Arnold~\cite{Arnold1965} as a case of three-dimensional
Euler flow, which could have chaotic trajectories in the Lagrangian sense.
M. Henon \cite{Henon1966} has provided a numerical evidence for chaos in the case
with $A=\sqrt{3}$, $B=\sqrt{2}$ and $C=1$. Independently, S. Childress~\cite{Childress1970} considered
the special case ($A = B = C = 1$) as a model for the kinematic dynamo effect.
X.H. Zhao et al.~\cite{Zhao1993} have studied the $ABC$-flow  by the method of
V. Melnikov \cite{Melnikov1963} to obtain
analytical criteria for the existence of chaotic streamlines and resonant streamlines in the $ABC$-flow.
D. Huang et al.~\cite{Huang1998} obtained an explicit analytical criterion for existence
of chaotic streamlines in the $ABC$-flow.
V. Arnold and E. Korkina~\cite{Arnold1983}, D. Galloway and U. Frisch~\cite{Galloway1984}, and
H. Moffatt and M. Proctor~\cite{Moffatt1985} performed numerical and analytical studies of
the dynamo action at finite values of the conductivity. They have shown that the $ABC$-flow can
excite a magnetic field. N. Brummell~\cite{Brummell2001} has investigated some linear and nonlinear
dynamo properties of the time-dependent $ABC$-flow.

T. Dombre et al.~\cite{Dombre1986} published in 1986 an extensive analysis of the $ABC$-flow
with any real values of the control parameters $A$, $B$, and $C$.
Their numerical  and analytical studies indicate that the $2:1$ resonances do not exist at
real values of the parameters, excepting for the special integrable
case with $A = B = 1$, and $C = 0$. In Ref.~\cite{Zhao1993} a theorem on the existence of a resonant
streamline near an elliptic point were proved.
\begin{figure*}[!ht]
\center{\includegraphics[width=0.8\textwidth]{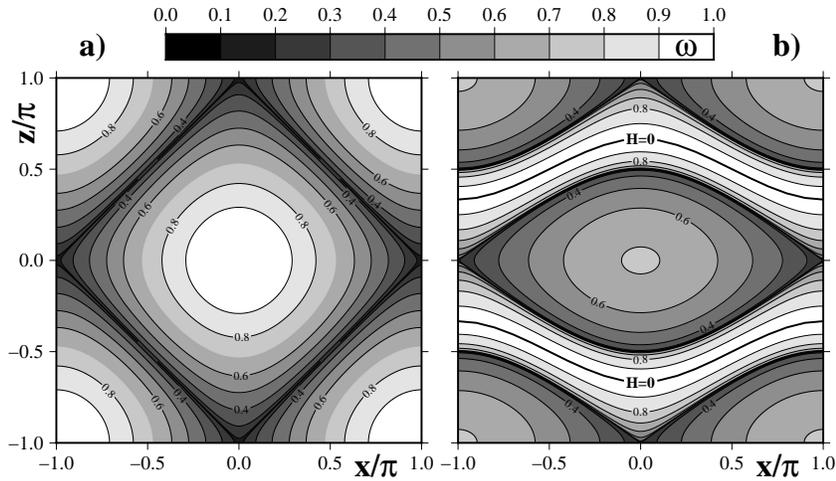}}
\caption{Unperturbed frequency ($\omega$) maps of the $ABC$-flow in the ($x$, $z$)-plane at a) $A=B=1$, $C=0$
and b)~$A=1$, $B=0.5$, $C=0$.}
\label{Fig1}
\end{figure*}

By using this theorem, one can obtain the same results as in Ref.~\cite{Dombre1986} for the
$n:1$ resonances, but our results are essentially different.
We obtain an explitic analytical criterion and numerical evidences
for existence of the $n:m$ resonances in the $ABC$-flow.
The point is that we prove existence of the largest $1:1$,
$2:1$ and $3:1$ resonances in regions of finite and infinite motion at any real values of
the parameters $A$ and $B$ and at sufficiently small $C$. We give analytical and numerical
evidences for existence of two branches of the $1:1$ and $2:1$ resonances in the region of finite motion and phase portraits of their bifurcations. The existence of the two resonance branches was completely missed in the abovementioned papers.

\section{The $ABC$-flow}

The $ABC$-flow has the following form
\begin{equation}
\label{int1}
\begin{aligned}
\dot{x} &= A\sin(z)+C \cos(y),\\
\dot{y} &= A\cos(z)+B \sin(x),\\
\dot{z} &= C\sin(y)+B \cos(x),
\end{aligned}
\end{equation}
where $x, y, z \in[-\pi,\pi)$. By using symmetry of the set (\ref{int1}), we can
let $1=A\geqslant B\geqslant C$.
Assuming that the elliptic point of a vortex is the initial point of reference frame
and using translation $x = x' +\pi/2$, equations (\ref{int1}) can  be  rewritten  as
\begin{equation}
\label{ABC1}
\begin{aligned}
\dot{x}&= \sin(z)+C \cos(y),\\
\dot{y} &= \cos(z)+B \cos(x),\\
\dot{z} &=  C\sin(y)-B \sin(x),
\end{aligned}
\end{equation}
where we omitted the prime over $x$.

In the case with $C=0$, the $ABC$-flow reduces to the two-dimensional set
\begin{equation}
\label{res1}
\dot{x}=-\frac{dH}{dz}= \sin(z),\quad
\dot{z}=\frac{dH}{dx}= -B \sin(x),
\end{equation}
where the streamfunction $$H=\dot y=\cos(z)+B \cos(x)=\text{const}$$ plays the role of a Hamiltonian.

In the case with $C=0$ and $B=1$, the two regions of finite motion are exist in the phase space.
Their properties are the same, but their streamfunctions $H$ are of opposite sign.
The stationary hyperbolic points of the 2D-set~(\ref{res1}) are also stationary points
of the 3D-set~(\ref{ABC1}) only in this case. In the case with $B\in(0,1)$, the region of infinite
motion appears with the size depending on values of the parameter $B$. The frequency map of
the set~(\ref{res1}) are shown in Fig.~\ref{Fig1} at some values of $B$.

The $ABC$-flow is symmetric, and we can consider the region with the streamfunction
$0\leqslant H\leqslant 1+B\equiv H_e$. We get $H_h\equiv1-B\leqslant H\leqslant 1+B\equiv H_e$
in the finite region, where $H_h$ and $H_e$~are values of the streamfunction $H$ at the hyperbolic
and elliptic points. In the infinite region
we will use values of the streamfunction $H$ in the interval $ 0\leqslant H\leqslant 1-B\equiv H_h$.

In Appendix~A we derive the expression for frequency of oscillations in the case of
the finite ($\omega_f$) and infinite ($\omega_i$) unperturbed motion
\begin{equation}
\label{wiwf}
\begin{gathered}
\omega_f(B, H)=\frac{\pi \sqrt{B}}{2 K(k_f)}, \qquad
\omega_i(B, H)=\frac{\pi \sqrt{B} }{k_i K(k_i)},\\
k^2_i(B, H)=\frac{1}{k_f^2(B, H)}=\frac{4B}{\left(1+B\right)^2-H^2},
\end{gathered}
\end{equation}
where $K(k)$~is the complete elliptic integral of the first kind, and $k$ is its modulus.

\section{Resonances in the case with $C \ll 1$}

In the case with $C\ll 1$  the set (\ref{ABC1}) can be represented as
\begin{equation}
\label{res1a}
\begin{gathered}
H= \cos(z)+B \cos(x),\\
\dot{x}=-\frac{dH}{dz}+C \cos(y),\quad
\dot{z}=\frac{dH}{dx}+C\sin(y).
\end{gathered}
\end{equation}
In this case $H\approx \text{const}$, so $y=Ht$ and $C\sin(y)$, $C\cos(y)$ can be considered
as a small periodic perturbation with the period $2\pi/H$.

The resonance condition implies that the following equation must be satisfied:
\begin{equation}
\label{res2}
n \omega(B, H) = m \nu,
\end{equation}
where $\omega$~---  the frequency of oscillations for an unperturbed motion,
$\nu=H$~--- the frequency of a periodic perturbation, $n$ and $m$~--- a pair of arbitrary
positive integers.

Therefore, the condition (\ref{res2}) can be rewritten in the finite region  as follows:
\begin{equation}
\label{res3a}
\frac{n \pi \sqrt{B}}{2 K(k_f)}=m H,
\end{equation}
and in the infinite region  as follows:
\begin{equation}
\label{res3b}
\frac{n \pi \sqrt{B} }{k_i K(k_i)}=m H.
\end{equation}

There is a solution for the streamfunction $H$ in infinite region at any values of $B$ and $m/n$
(see Appendix~B).

Finite region is restricted to the values of $H\in[H_h,H_e]$.
The function $w_f(H)$ is a concave (see Appendix~B) and $w_f(H_h)=0$, $w_f(H_e)>0$.
So we can assume that two restricted lines for the region of solutions of eq.~(\ref{res3b})
exist, and this equation has no more than two solutions.
The first line, $f_1H$, passes through
the point $(0,0)$ and is tangent to the curve $w_f(H)$. The value of streamfunction at the
tangent point, $H_\text{tan}$, can be computed from equation
\begin{equation}
\label{res5}
\left.{\frac { d\omega_f}{dH}}\right\vert_{H=H_\text{tan}}H_\text{tan}=\omega_f(H_\text{tan}),
\end{equation}
or, in the expanded form as
\begin{equation}
\label{res7}
\frac{H_\text{tan}^2\left[E(k_f)-(1-k_f^2)K(k_f)\right]}{\left((1+B)^2-H_\text{tan}^2\right)(1-k_f^2)K(k_f)}=1,
\end{equation}
where $E(k_f)$ is the complete elliptic integral of the second kind.
The slope coefficient $f_1$ depends on the parameter $B$ and can be obtained as
\begin{equation}
\label{res6}
f_1(B)=\left.{\frac { d\omega_f}{dH}}\right\vert_{H=H_\text{tan}}\!=\frac{\pi H_\text{tan}}{8\sqrt{B}}\frac{\left[E(k_f)-(1-k_f^2)K(k_f)\right]}{k_f^2(1-k_f^2)K^2(k_f)}.
\end{equation}
The second line, $f_2H$, connects the point $(0,0)$ with the final point of curve $w_f(H)$,
$(H_e,w_f(H_e))$. The slope coefficient $f_2$ also depends on parameter value $B$ and can be found as
\begin{equation}
\label{res8}
f_2(B)=\frac{\omega_f(H_e)}{H_e}=\frac{\sqrt{B}}{1+B}.
\end{equation}

\begin{figure}[!ht]
\centerline{\includegraphics[width=1\linewidth]{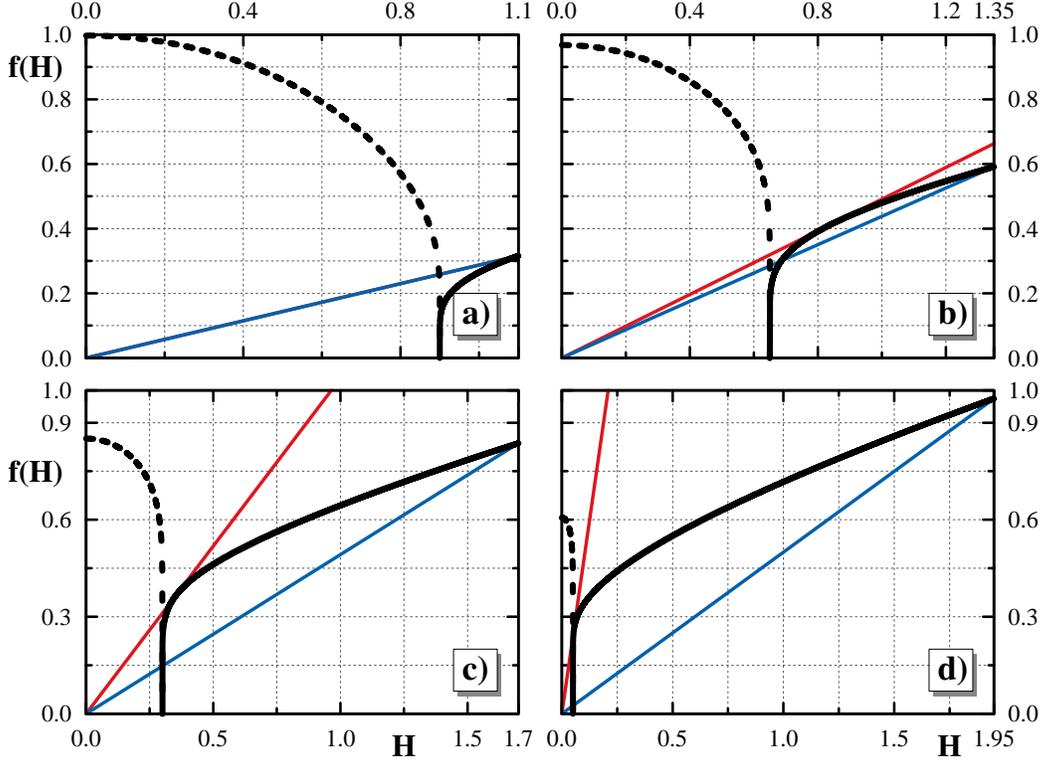}}
\caption{Graphical solutions of equations (\ref{res3a}) and (\ref{res3b}) are shown
at a)~$B=0.1$, b)~$B=0.35$, c)~$B=0.7$, and d)~$B=0.95$. The black solid line is the function $\omega_f(H)$, the black dash line~--- $\omega_i(H)$,
the solid red line~--- $f_1H$, and the solid blue line~--- $f_2H$.}
\label{Fig2}
\end{figure}

\begin{figure}[!ht]
\centerline{\includegraphics[width=1\linewidth]{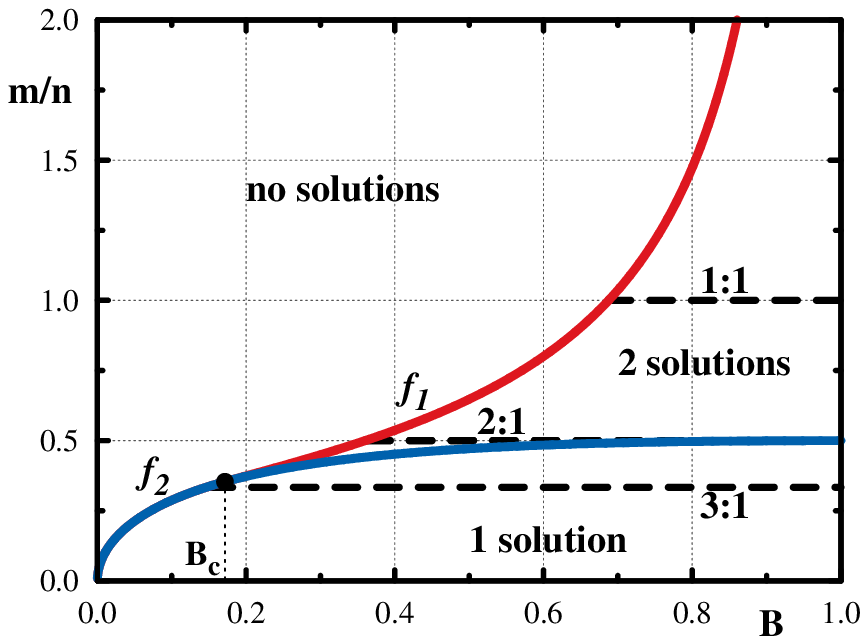}}
\caption{Graphical solution of eq. (\ref{res3a}) in the parametric space.
The largest resonances $1:1$, $2:1$, and $3:1$ are shown by the dash lines.}
\label{Fig3}
\end{figure}

Equation~(\ref{res7}) has a solution only if $B\geqslant B_c\approx 0.1716$. The critical value
of the parameter $B$ can be found from equation
\begin{equation}
f_1(B_c)=f_2(B_c).
\end{equation}
Graphical solutions of eqs.~(\ref{res3a}) and~(\ref{res3b}) are shown on Fig.~\ref{Fig2}.

Thus, the number of solutions of eq.~(\ref{res3a}) in the finite region depends on the parameter
value $B$ and of the order of the resonance $n:m$.
\begin{itemize}
\item $B\leqslant B_c$, $m/n>f_2(B)$. Eq.~(\ref{res3a}) has no solutions, and there are no resonances
of the order $n:m$.
\item $B\leqslant B_c$, $m/n\leqslant f_2(B)$. Eq.~(\ref{res3a}) has one solution, and
there is one resonance of the order $n:m$.
\item $B>B_c$, $m/n>f_1(B)$. Eq.~(\ref{res3a}) has no solutions, and there are no resonances
of the order $n:m$.
\item $B>B_c$, $f_2(B)<m/n\leqslant f_1(B)$. Eq.~(\ref{res3a}) has two solutions, and there are
two resonances of the order $n:m$.
\item $B>B_c$, $m/n\leqslant f_2(B)$. Eq.~(\ref{res3a}) has one solution, and there is one resonance
of order the $n:m$.
\end{itemize}
All these solutions are graphically shown in Fig.~\ref{Fig3}.

\begin{figure}[!ht]
\center{\includegraphics[width=1\linewidth]{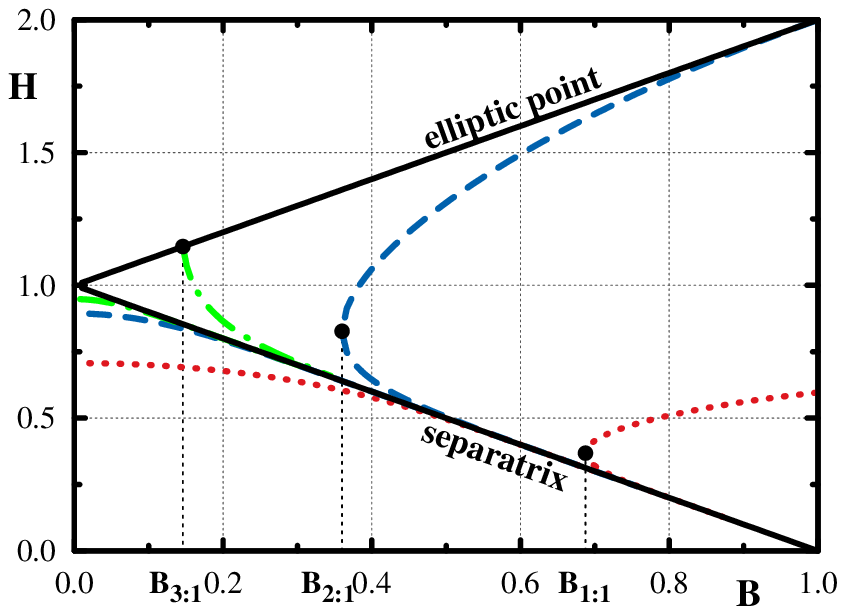}}
\caption{Energy of the resonances defined by (\ref{res3a}) and (\ref{res3b}): the dot lines~--- $1:1$
resonances; the dash lines~--- $2:1$ resonances;  the dot-dash lines~--- $3:1$ resonances.
The energies between separatrix and the elliptic point correspond to the finite region, the energies
below the separatrix line correspond to the infinite region.}
\label{Fig4}
\end{figure}

\begin{figure}[!ht]
\center{\includegraphics[width=1\linewidth]{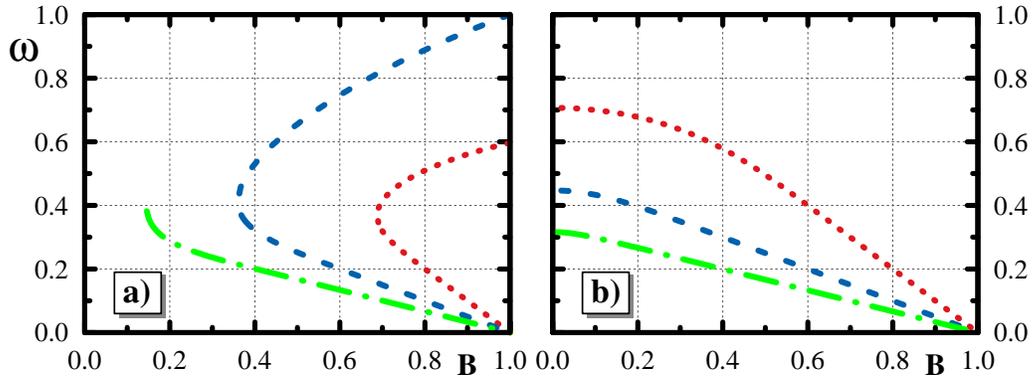}}
\caption{Frequency for motion at a resonance trajectory in a) the finite region
and b) the infinite region.
The dot lines~--- $1:1$ resonances; the dash lines~--- $2:1$ resonances;  the dot-dash lines~--- $3:1$
resonances.}
\label{Fig5}
\end{figure}

Equations~(\ref{res3a}) and (\ref{res3b}) can be solved numericalally. Locations of the $1:1$, $2:1$,
and $3:1$ resonances in the finite and infinite regions are presented in Fig.~\ref{Fig4}.
The $1:1$ resonance in the finite region exists in the case $B\geqslant B_{1:1}\approx 0.6874$.
The $2:1$ resonance in the finite region exists in the case $B\geqslant B_{2:1}\approx 0.3600$.
The $3:1$ resonance in the finite region exists in the case $B\geqslant B_{3:1}\approx 0.1459$.
Figure~\ref{Fig5} shows the frequency for trajectories of the $1:1$, $2:1$, and $3:1$ resonances.

\begin{figure*}[!ht]
\center{\includegraphics[width=0.85\textwidth]{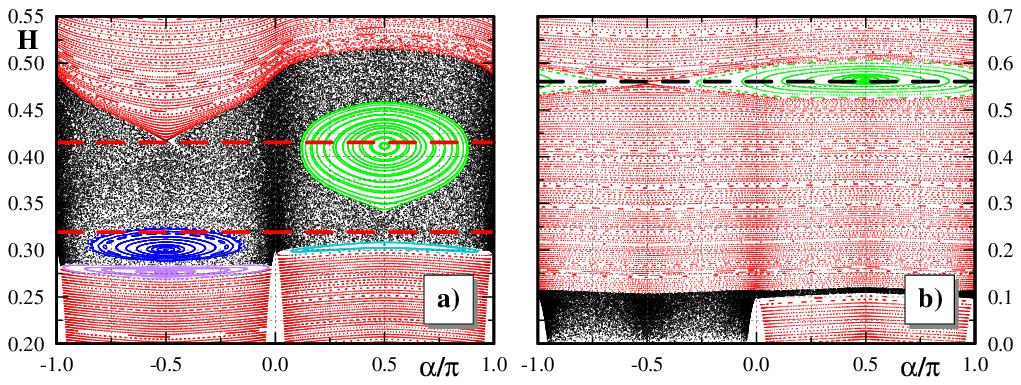}}
\caption{Poincare sections in the $\alpha$\,--\,$H$ space. The dash lines~--- theoretical values of energy for two branches of the $1:1$ resonance. a) $B=0.7$ and $C=0.005$; b) $B=0.9$ and $C=0.005$.}
\label{n1}
\end{figure*}

\section{Numerical evidences for the $n:m$ resonances}
In this section we  consider the results of numerical experiments.
Poincare sections are given in the space $\alpha$\,--\,$H$, where $\alpha$ is the angle defined by the expression $\tan\alpha=z/x$, and $H$ is the energy given in~\ref{res1a}.

Figure~\ref{n1}a demonstrates the Poincare section with all the $1:1$ resonances in the region
$H>0$.
The ``purple'' and ``dark turquoise'' islands are resonances $1:1$ in two regions of the infinite
motion above ($z>0$) and below ($z<0$) the separatrix (see Fig.~\ref{Fig1}). The ``blue'' island
is the ``lower'' resonance branch of the $1:1$ resonance in the region of finite motion, the ``green''~--- the ``upper'' branch of the same resonance (see Fig.~\ref{Fig4}). The dashed line shows
a theoretical value of $H$ given by Eq.~\ref{res3a} for these resonances in the case of sufficiently
small $C$.
The apparent deviation of the ``blue'' resonant island from the theoretical line is due to
a proximity of the separatrix.
\begin{figure}[!ht]
\center{\includegraphics[width=0.69\linewidth]{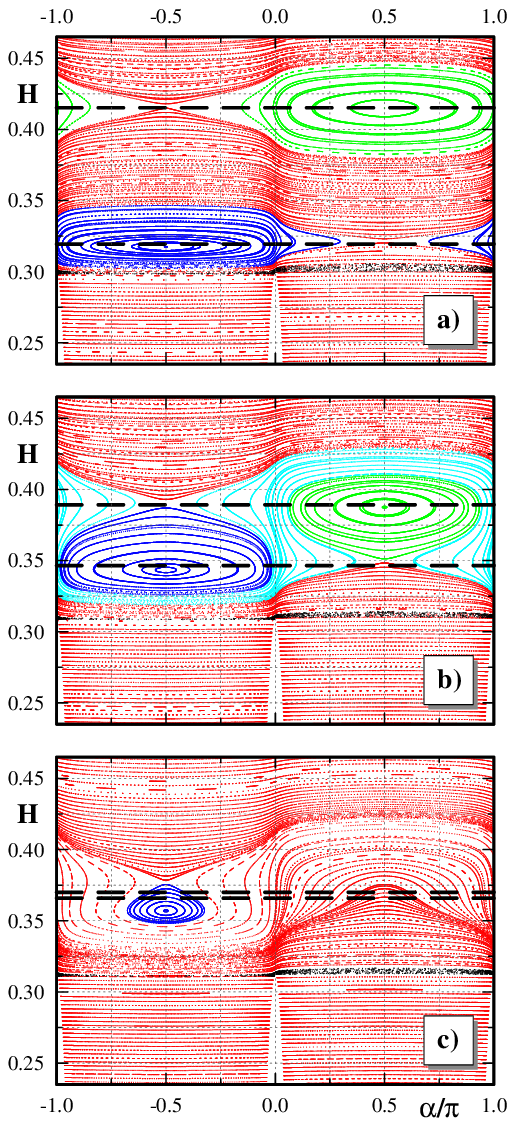}}
\caption{Poincare sections in the $\alpha$\,--\,$H$ space.
The black dash lines~--- theoretical values of energy for the resonance $1:1$.
a) $B=0.7$ and $C=0.0005$; b) $B=0.69$ and $C=0.0005$;
c) $B=0.6875$ and $C=0.0005$.}
\label{n2}
\end{figure}
\begin{figure}[!ht]
\center{\includegraphics[width=0.59\linewidth]{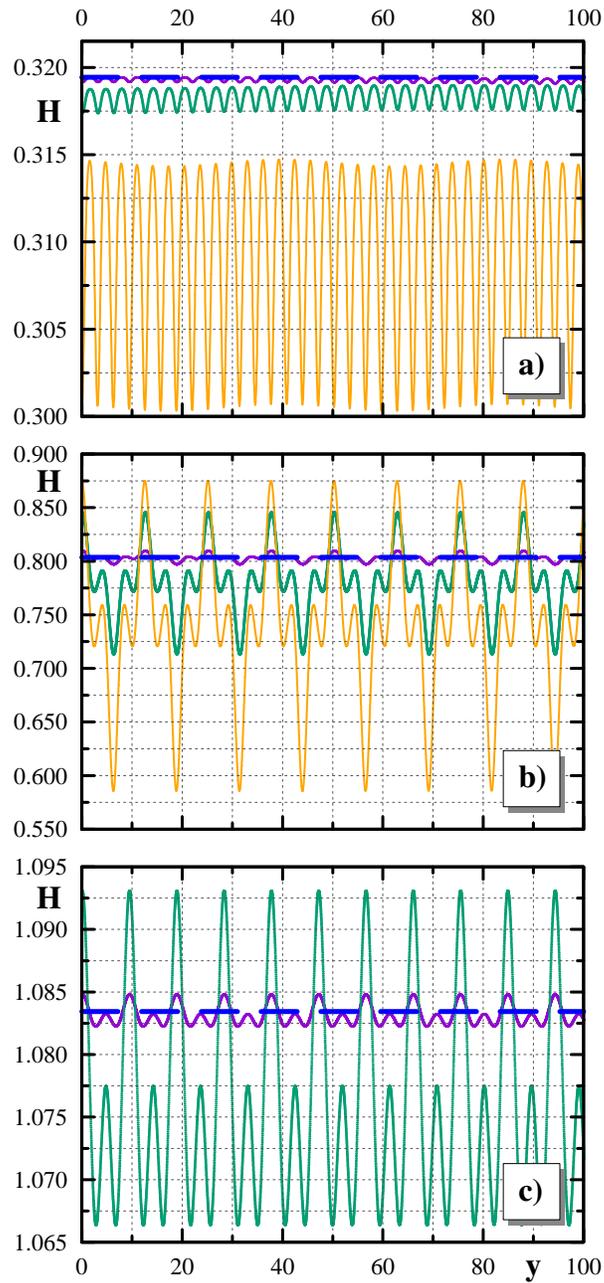}}
\caption{Value of energy for a trajectory near elliptic points of the resonances.
The blue dashed line~--- a theoretical value of energy of the corresponding resonance.
a) The resonance $1:1$ proximal to the separatrix (``blue`` island), $B=0.7$, the purple curve~--- $C=0.0001$, the green curve~--- $C=0.0005$, and
the yellow curve~--- $C=0.005$.
b) The resonance $2:1$ proximal to the separatrix (``blue'' and ``orange'' islands), $B=0.3605$, the purple curve~--- $C=0.005$,
the green curve~--- $C=0.05$ and the yellow curve~--- $C=0.1$.
c) The resonance $3:1$, $B=0.15$, the purple curve~--- $C=0.005$ and the green curve~--- $C=0.05$.}
\label{n4}
\end{figure}

\begin{figure*}[!ht]
\center{\includegraphics[width=0.9\textwidth]{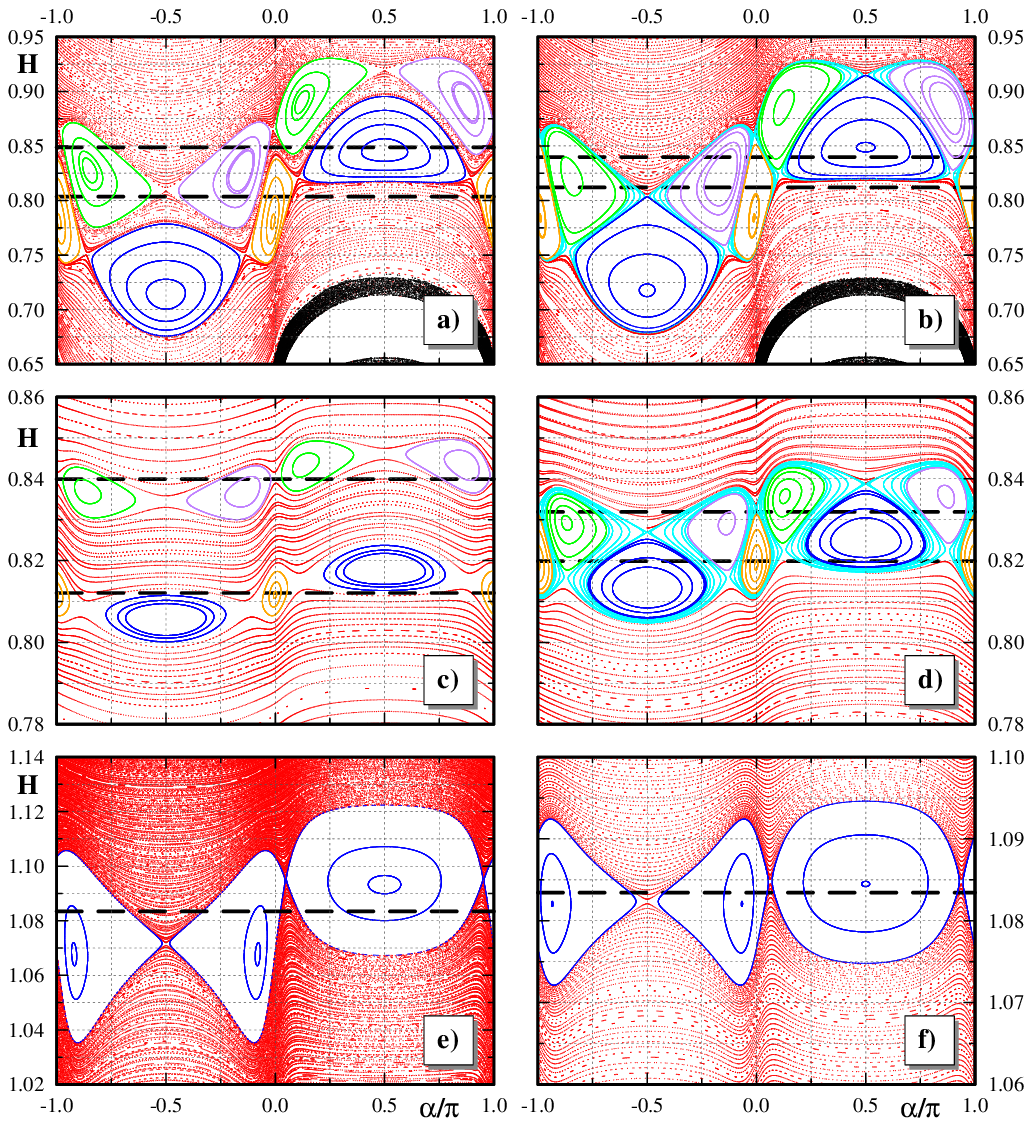}}
\caption{Poincare sections in $\alpha$\,--\,$H$ space. Black dash line~---
theoretical value of energy for the corresponding resonance.
a) Resonance $2:1$, $B=0.3605$, $C=0.05$;
b) $2:1$, $B=0.3602$, $C=0.05$;
c) $2:1$, $B=0.3602$, $C=0.005$;
d) $2:1$, $B=0.36005$, $C=0.005$;
e) $3:1$, $B=0.15$, $C=0.05$;
f) $3:1$, $B=0.15$, $C=0.005$.}
\label{n3}
\end{figure*}

In the case of $B=0.9$, the ``lower'' branch of the resonance $1:1$ merges with the separatrix and
the corresponding ``blue'' island collapses (see Fig.~\ref{n1}b).
Figure~\ref{n2} shows merging and disappearance of the two resonance branches $1:1$ as the parameter
$B$ decreases. Initially, the two chains of islands are separated from each other by regular phase
trajectories (Fig.~\ref{n2}a). As the parameter $B$ decreases, the chains approach, and their
separatrices are reconnected from a heteroclinic topology to a homoclinic one (see Fig.~\ref{n2}b).
The value of $B$, at which reconnection occurs, depends on the size of the islands and is determined
by the perturbation value of $C$. With a further decrease in $B$, the elliptic point of one chain
merges with the hyperbolic point of another chain, and the island disappears. This merging occurs
at different values of $B$ for each pair of points, as can be seen in Fig.~\ref{n2}c, where the right
pair of points has already merged, but the elliptic point of the ``lower'' chain and the hyperbolic
point of the ``upper'' chain still exist. The reconnection of separatrices and the merge of resonance chains resemble similar bifurcations in other degenerate systems, described, for example, in \cite{Simo98, Morozov2014, Budyansky2009}.

The resonance $2:1$ has a complicated structure. Each of its chain consists of two pairs of islands
clearly seen in Fig.~\ref{n3}a--d, where each pair has the own color. Behavior of these chains as
$B$ varies is analogous to behavior of the $1:1$ resonance chains. With decreasing $B$,
the separatrices of two chains are reconnected (Fig.~\ref{n3}b), but as the perturbation $C$
decreases, the chains can split again (Fig.~\ref{n3}c). However, a further decrease in $B$ leads
to another reconnection (Fig.~\ref{n3}d): a pairwise merging of elliptic and hyperbolic points
and disappearance of the resonance.

The resonance $3:1$ ($m/n=1/3$) lies below the critical value $f_1(B_c)=f_2(B_c)\approx 0.3536$,
therefore it has only one chain of islands. We show in Figs.~\ref{n3}e and~f some possibilities
for this resonance with different values of $C$. As the parameter $B$ decreases, the resonance
disappears due to approach to the elliptical point of the unperturbed system and its merging with
this point.
If the value of perturbation $C$ is nonzero, the energy of the elliptic point can deviate from the
theoretically calculated one. However, as $C$ tends to zero, the energy of the elliptic point tends
to the energy calculated from Eq.~(\ref{res3a}). Figure~\ref{n4} shows the dependence of the energy
of a trajectory close to an elliptic point on the $y$ axis for the resonances $1:1$ (Fig.~\ref{n4}a),
$2:1$ (Fig.~\ref{n4}b), and $3:1$ (Fig.~\ref{n4}c) at different values of $C$. For all the three
resonances,
the average energy decreases with increasing $C$, that is, the resonances shift toward the separatrix.

\section{Conclusion}

The stochastic layer has a strong influence on generation of the magnetic field. It is believed
that stretching of material lines in a flow due to dynamical chaos leads to the so-called
``fast dynamo'' regime, in which the magnetic field grows exponentially fast. Resonances have a rather serious effect on the stochastic layer, as it can be expanded by formation of an additional
stochastic layer on the resonance separatrix, or it can weaken its influence due to sticking of
trajectories to resonance islands.
If one knows where the resonances located in the phase space, it becomes possible to
control them with the help of an additional small time-dependent perturbation, for example, to destroy them and increase the total
volume of the stochastic layer.

Being motivated by this fact,
we have studied the resonances in the $ABC$-flow in the near integrable case ($C\ll 1$).
The analytical conditions (\ref{res3a}) and (\ref{res3b}) for the $n:m$ resonances have been obtained. It was shown numerically that the resonances $1:1$, $2:1$, and $3:1$ exist, and their energies are equal to theoretical energies in the near integrable case.
We provided analytical and numerical evidences for existence of two branches of the $1:1$ and $2:1$ resonances in the region of finite motion. It is interesting that the existence of the two branches of those resonances is not accompanied by a degeneracy in the system. This is due to a peculiarity
of the resonance condition, in which the frequency of perturbation depends on the energy.
Two branches of the corresponding resonance can interact with each other, which is accompanied by
a reconnection of their separatrices and a destruction (creation) of pairs of elliptic and hyperbolic points.

\section*{Acknowledgments}
The authors would like to thank Prof. S. Prants for a critical reading the manuscript and valuable
comments. This work has been supported by the Russian Science Foundation (project
no.~16--17--10025).

\appendix

\section{The frequency of unperturbed motion in finite and infinite regions of motion}
Let us find the frequency for finite trajectories $\omega_f$.
The frequency of oscillations is defined as
\begin{equation}
\label{fq1}
\begin{aligned}
\omega=\frac{dH}{dI}=\left[\frac{dI}{dH}\right]^{-1},
\end{aligned}
\end{equation}
where $I$~is the action which can be found by using the following expression
from Ref.~\cite{Zaslavsky1991}:
\begin{multline}
\label{fq2}
I=\frac{1}{2\pi} \,\oint z dx =\\
\frac{1}{\pi} \,\int\limits_{-\arccos\left(\frac{H-1}{B}\right)}^
{\arccos\left(\frac{H-1}{B}\right)}\arccos(H-B\cos(x))dx=\\
\frac{2}{\pi} \,\int\limits_{0}^{\arccos\left(\frac{H-1}{B}\right)}
\arccos(H-B\cos(x))dx,
\end{multline}
where $H_h\equiv 1-B < H < 1+B\equiv H_e$ and $0\leqslant B\leqslant 1$.
Substituting (\ref{fq2}) to (\ref{fq1}), we get
\begin{multline}
\label{fq3}
\scalebox{0.9}{$\displaystyle \omega_f=\left[\frac{d}{dH}\left(\frac{2}{\pi}\!\!\!\!\!\int\limits_{0}^{\arccos\left(\frac{H-1}{B}\right)}\!\!\!\!\!\!\!
\arccos(H-B\cos(x))dx\right)\right]^{-1}$}\!\!\!\!\!\!=\\
\scalebox{0.9}{$\displaystyle\left[\frac{2}{\pi}\left(-\!\!\!\!\!\!\!\int\limits_{0}^{\arccos\left(\frac{H-1}{B}\right)}\!\!\!\!\!\!\!\!\!
\frac{dx}{\sqrt{1-\left[H+B\cos(x)\right]^2}}-
\frac{\arccos(1)}{B\sqrt{1-\left[\frac{H+1}{B}\right]^2}}\right)\right]^{-1}$}\!\!\!.
\end{multline}
After the replacement $\xi=\cos(x)$, we get
\begin{equation}
\label{fq4}
\begin{aligned}
\omega_f=\left[-\frac{2}{\pi B}\int\limits_{\frac{H-1}{B}}^{1}
\frac{d\xi}{\sqrt{1-\xi^2}\,\sqrt{\frac{1}{B^2}-\left[\frac{H}{B}+
\xi\right]^2}}\right]^{-1}.\\
\end{aligned}
\end{equation}
Now we represent (\ref{fq4}) as an elliptic integral by using $\xi=\xi(\phi)$ from~\cite{Korn} and get
\begin{equation}
\label{fq5}
\begin{aligned}
\frac{d\xi}{\sqrt{1-\xi^2}\,\sqrt{\frac{1}{B^2}-\left[\frac{H}{B}+\xi\right]^2}}=
\mu\frac{d\phi}{\sqrt{1-k_f^2\sin^2(\phi)}}
\end{aligned}
\end{equation}
to obtain
\begin{equation}
\label{fq6}
\begin{aligned}
\omega_f=\left[-\frac{2}{\pi B}\int\limits_{0}^{\pi/2}
\mu\frac{d\phi}{\sqrt{1-k_f^2\sin^2(\phi)}}\right]^{-1},
\end{aligned}
\end{equation}
where $\mu=\sqrt{B}$ and $k_f^2=\frac{\left(B+1\right)^2-H^2}{4B}$.
Thus, the frequency along finite trajectories is
\begin{equation}
\label{fq7}
\begin{aligned}
\omega_f=\frac{\pi \sqrt{B}}{2 K(k_f)},
\end{aligned}
\end{equation}
where we omit the sign.

Let us find frequency for infinite trajectories $\omega_i$. In this case the action $I$ is defined as
\begin{multline}
\label{fq8}
I= \frac{1}{2\pi} \,\int\limits_{-\pi}^
{\pi}\arccos(H-B\cos(x))dx=\\
\frac{1}{\pi} \,\int\limits_{0}^{\pi}
\arccos(H-B\cos(x))dx,
\end{multline}
where $ 0< H < 1-B\equiv H_h$.\\
\begin{multline}
\label{fq9}
\omega_i=\left[\frac{d}{dH}\left(\frac{1}{\pi} \,\int\limits_{0}^{\pi}
\arccos(H-B\cos(x))dx\right)\right]^{-1}=\\
\left[-\frac{1}{\pi}\int\limits_{0}^{\pi}
\frac{dx}{\sqrt{1-\left[H-B\cos(x)\right]^2}}\right]^{-1}.\\
\end{multline}
After some replacements~\cite{Korn}, we get
\begin{equation}
\label{fq10}
\begin{aligned}
\omega_i=\left[-\frac{1}{\pi B}\int\limits_{0}^{\pi/2}
\lambda\frac{d\phi}{\sqrt{1-k^2_i\sin^2(\phi)}}\right]^{-1},
\end{aligned}
\end{equation}
where $\lambda=\frac{2B}{\sqrt{(B+1)^2-H^2}}$ and $k^2_i=1/k_f^2=
\frac{4B}{\left(B+1\right)^2-H^2}$. So, the frequency for infinite trajectories can be written as
\begin{equation}
\label{fq11}
\begin{aligned}
\omega_i=\frac{\pi \sqrt{B}}{k_i K(k_i)},
\end{aligned}
\end{equation}
where we also omit the sign.

\section{A note on the number of solutions for the resonance conditions (\ref{res3a})
and (\ref{res3b})}

Let us consider the infinite region. At $H\in[0, H_h]$, the frequency $w_i$ is a monotonically
decreasing function since
\begin{equation}
\label{N3}
\frac{dw_i}{dH}=-\frac{\pi H\sqrt{(1+B)^2-H^2}E(k_i)}{2\left[(-H_h)^2-H^2)\right]K^2(k_i)}<0,
\end{equation}
and the straight line $(m/n) H$ is a monotonically increasing function.
At the edges, the modulus $k_i$ is written as
\begin{equation}
\label{N1}
k_i^2(B, 0)=\frac{4B}{(1+B)^2}, \qquad
k_i^2(B, H_h)=1.
\end{equation}
The frequency $w_i$ for the modulus $k_i$ and the straight line $(m/n) H$ have the following
form at the edges:
\begin{equation}
\label{N2}
\begin{aligned}
w_i(B, 0)=\frac{\pi(1+B)}{2K\left(k_i(B, 0)\right)}&\geqslant 0,\\
w_i(B, H_h)=0&<(m/n)H_h.
\end{aligned}
\end{equation}
Thus, we obtain that one and only one solution of eq.~(\ref{res3b}) exists at any values of $m/n$ and~$B$.

Now we face the finite region where $H\in[H_h, H_e]$. To prove that $w_f(H)$ is a concave, we need
to consider $d^2w_f/dH^2$. If $d^2w_f/dH^2\leqslant 0$ in the interval $H\in[H_h, H_e]$, then
we can say that $w_f(H)$ is concave.
The first derivative
\begin{equation}
\label{N4}
\frac{dw_f}{dH}=\frac{\pi H \left[ E(k_f)-(1-k_f^2)K(k_f)\right]}{8\sqrt{B}
k_f^2(1-k_f^2) K^2(k_f)}
\end{equation}
is always positive, therefore $w_f(H)$ is a monotonically increasing function.

The second derivative can be written as
\begin{multline}
\label{N5}
\frac{d^2w_f}{dH^2}=\\
\frac{\pi H^2  \left[ E(k_f)-K(k_f)\right] \left[2E(k_f)+(k_f^2-1)K(k_f)\right]}{32B^{3/2}k_f^4(k_f^2-1)^2 K^3(k_f)}-\\
\frac{\pi (B+1)^2(k_f^2-1)K(k_f)\left[E(k_f)+(k_f^2-1)K(k_f)\right]}{32B^{3/2}k_f^4(k_f^2-1)^2 K^3(k_f)}.
\end{multline}
Since $H^2=(B+1)^2-4Bk_f^2$, then $d^2w_f/dH^2$ is a function of two variables, $k_f\in[0,1]$
and $B\in(0,1]$. Unfortunately, we were able to show only numerically that this function is negative
throughout the domain of its definition.

\end{document}